\UseRawInputEncoding
\documentclass[aps,prl,reprint,floatfix,amsmath,amssymb]{revtex4-1}

\usepackage{epsfig}
\usepackage{graphicx}
\usepackage{dcolumn}
\usepackage{bm}
\usepackage[colorlinks=true,citecolor=blue, dvipdfm]{hyperref}
\usepackage{amssymb}

\begin{document}
\title{Nucleation and growth manifest universal scaling, surely}

\author{Fan Zhong}
\affiliation{School of Physics and State Key Laboratory of Optoelectronic Materials and
Technologies, Sun Yat-sen University, Guangzhou 510275, People's Republic of China}

\date{\today}

\begin{abstract}
When a system is brought to a metastable state, nuclei of the equilibrium phase form and grow. This is the well-known nucleation and growth of first-order phase transitions. Near a critical point of a continuous phase transition, critical phenomena such as critical opalescence characterized by universal scaling emerge. These two sets of behavior are so completely different that it might appear absurd to ask whether nucleation and growth exhibit even a slight universal scaling. Here we show that universal scaling is indeed manifested in standard nucleation and growth by introducing a non-universal initial time to a recently proposed method. This generalizes the method and provides a different perspective for studying universal scaling behavior and hence universality classes of nucleation and growth.
\end{abstract}

\maketitle

When clean water is heated to above $100$ degree Celsius at the ambient pressure, bubbles form and expand in the bulk water. This is nucleation and growth of the vapor phase in the liquid phase~\cite{Becker,Becker1,Becker2,Zeldovich,Avrami,Avrami1,Avrami2,Langer67,books,books1,books2,Oxtoby92,Oxtoby921,Oxtoby922,Oxtoby923,Binder16,zhong18}, a standard mode of first-order phase transitions (FOPTs)~\cite{Gunton83,Binder,Binder2}. At an elevated temperature and pressure, the critical point, water becomes milky white, showing critical opalescence~\cite{Stanley}. This is a generic critical phenomenon exhibiting in continuous phase transitions that are characterized by universal scaling behavior~\cite{Mask,Cardyb,Justin,Amit}. The two sets of behavior are so completely different that it might appear absurd to ask whether nucleation and growth exhibit even a slight universal scaling. Here, we will offer a positive answer to the question for the standard nucleation and growth of a system evolving from its metastable state by introducing a non-universal initial time to a recently proposed method for the case of driving (via varying an external field) a system from one equilibrium state to another~\cite{zhong24}. This generalizes the method and provides a different perspective for studying universal scaling behavior and hence universality classes of nucleation and growth and may thus shed light on the long standing issue of the large discrepancy between theoretical predictions and experiments in nucleation and growth~\cite{Oxtoby92,Oxtoby921,Oxtoby922,Oxtoby923,Filion,Filion1,Filion2}.

Consider a conventional Ginzburg-Landau functional~\cite{Mask,Cardyb,Justin,Amit,zhong24,zhongl05,zhong16,zhong12}
\begin{equation}
F_4(\phi)=\int d^dx\left\{f_4(\phi)+\frac{1}{2}\left(\nabla\phi\right)^2\right\}\label{gl}
\end{equation}
in $d$-dimensional space with a quartic free energy
\begin{equation}
f_4(\phi)=\frac{1}{2}a_2\phi^2+\frac{1}{4}a_4\phi^4-H\phi\label{f4}
\end{equation}
for an order parameter $\phi$ and its ordering field $H$, where $a_2$ is a reduced temperature and $a_4>0$ a coupling constant. Dynamics can be studied by the usual Langevin equation
\begin{equation}
\frac{\partial\phi}{\partial t}=-\lambda\frac{\delta F_4}{\delta\phi}+\zeta=-\lambda\left(a_2\phi+a_4\phi^3-H-\nabla^2\phi\right)+\zeta,\label{lang2d}
\end{equation}
where $\lambda$ is a kinetic constant and $\zeta$ a Gaussian white noise whose averaged moments satisfying
$\langle\zeta({\bf x},t)\zeta({\bf x}',t')\rangle=2\lambda\sigma\delta(t-t')\delta({\bf x}-{\bf x}')$.
with a noise amplitude $\sigma$. The model possesses a continuous transition at the critical point $a_2=0$ and $H=0$ but FOPTs for all $a_2<0$ between two ordered phases with $\phi_{\rm eq}=\pm\sqrt{-a_2/a_4}=\pm M_{\rm eq}$ at $H=0$ in the mean field (MF) approximation in which $\phi$ is spatially uniform. In this approximation, the FOPTs can only take place beyond the so-called MF spinodal point ($H_{s {\rm MF}}$, $M_{s {\rm MF}}$), at which the free energy barrier between the two phases vanishes and the system becomes unstable. This prompts us to let~\cite{zhongl05,zhong16}
$\phi=M_s+\varphi$,
one finds $f_4(\phi)=f_4[M_s]+f_3(\varphi)+a_4\varphi^4/4$ with
\begin{equation}
f_3(\varphi)=\frac{1}{2}\tau\varphi^2+\frac{1}{3}a_3\varphi^3-h\varphi,\label{f3}
\end{equation}
where
\begin{equation}
\tau=a_2+3a_4M_s^2,\qquad \hat{H}_s=a_2M_s+a_4M_s^3,\label{tauh}
\end{equation}
$h=H-H_s$ with $H_s=\hat{H}_s$, and $a_3=3a_4M_s$. $\tau$ and $h$ are an effective reduced temperature and an effective field, respectively, that become zero exactly at $M_s=M_{s {\rm MF}}=\pm\sqrt{-a_2/3a_4}$, $H_s=H_{s {\rm MF}}=2a_2M_{s {\rm MF}}/3$.
One sees therefore that $\tau=0$ and $h=0$ exactly at the MF spinodal point, similar to $a_2=0$ and $H=0$ at the critical point. Accordingly, we will see that the transitions are in fact controlled by the effective cubic theory $F_3$ ($F_4$ with $f_4$ replaced by $f_3$ and $\phi$ by $\varphi$) near the spinodal where the quartic term is negligible instead of the original quartic theory.

We have used $H_s$ and $M_s$ to represent the general spinodal point. Accordingly, there will be a fluctuation contribution $\delta h$, Eq.~(\ref{dth}) below, to $H_s$ such that $H_s=\delta h+\hat{H}_s$. One might argue that sharp spinodal points do no exist for systems with short-range interactions, since they depend on coarse-graining scales~\cite{Gunton83,Binder16,Binder2}. However, for the zero-dimensional Langevin equation, Eq.~(\ref{lang2d}) in the absence of the gradient term, the transitions between the two equilibrium phases can occur before the MF spinodal points are reached~\cite{zhong24}. This may also be regarded as the absence of the real spinodal points since such points may be considered to depend on $\sigma$. However, it has been clearly demonstrated that such a case is again well described by the cubic theory~\cite{zhong24}. So is the two-dimensional case~\cite{zhong24}. These show that spinodal points are essential even in these cases. A possible method to estimate them may be to match the relevant time scales, such as the escape or nucleation and growth time scale and the driving time scale~\cite{zhong18}. However, besides assuming their existence in some forms, we will not need to know them below~\cite{zhong24}.

Upon considering fluctuations, to one-loop order, the critical point has a displacement,
\begin{equation}
\delta a_2\propto a_4\int \frac{d^dk}{a_2+k^2},\label{da2}
\end{equation}
at $H=0$ from the quartic theory $F_4$~\cite{Mask,Cardyb,Justin,Amit}, where $k$ is a wave number and dimensionless constants have been ignored. Universal critical behavior is found from the renormalization-group theory~\cite{Mask,Cardyb,Justin,Amit}. Similarly, the MF transition point $(\tau,h)=(0,0)$ has finite fluctuation contributions,
\begin{equation}
\delta\tau\propto a_3^2\int \frac{d^dk}{(\tau+k^2)^2},\qquad \delta h\propto a_3\int \frac{d^dk}{\tau+k^2},\label{dth}
\end{equation}
in the cubic theory $F_3$~\cite{zhong16}. Also, the renormalization-group theory dictates that the universal long-wavelength long-time behavior is given by~\cite{zhongl05,zhong16}
\begin{equation}
m=b^{-\beta/\nu}g(\tau b^{1/\nu}, hb^{\beta\delta/\nu}, tb^{-z},a_3^*b^{\epsilon/2}),\label{mheq}
\end{equation}
controlled by a nontrivial fixed point $a_3^{*2}=-\epsilon b^{-\epsilon}/6N_d$, where $m=\langle\varphi\rangle=\langle\phi\rangle-M_s=M-M_s$, $\epsilon=6-d$, $N_d$ is a $d$-dependent constant, $b$ a length scaling factor, $g$ a universal scaling function, and $\beta$, $\delta$, $\nu$, and $z$ are the cubic counterparts of the critical exponents with identical symbols and meaning~\cite{zhongl05,zhong16}. In the MF theory, $z=2$, $\nu=1/2$, $\beta=\nu(d-2)/2=1$, and $\delta=\nu(d+2)/2=2$ for $d=6$, the upper critical dimension of the cubic theory~\cite{zhongl05,zhong16}. In Eq.~(\ref{mheq}), $\tau$ now denotes the fluctuation-displaced effective temperature similar to its critical counterpart~\cite{Mask,Cardyb,Justin,Amit}.

Equation~(\ref{mheq}) is identical with its critical counterpart except for the finite $M_s$ and $H_s$. More importantly, the fixed point $a_3^*$ is imaginary
and is thus usually considered to be unphysical, though the exponents are real~\cite{zhongl05,zhong16}. Yet, counter-intuitively, it has been shown that imaginariness is physical in order for the theory to be mathematically convergent, since upon renormalization the degrees of freedom that need finite free energy costs for nucleation are integrated away, leaving behind a truly unstable system at the transition point. As a result, analytical continuation is necessary and nucleation is irrelevant to scaling~\cite{zhong12}. This indicates that nucleation and growth ought to exhibit universal scaling.

For real transitions, the above core result must be supplemented with additional ingredients~\cite{zhong24}. The first one is the quartic term, since the FOPTs arise from the original quartic theory whereas the cubic theory is only effective. Because $a_3=3a_4M_s$, $a_4$ scales as $a_4b^{[a_4]}$ with the scale dimension (denoted by square brackets) $[a_4]=[a_3]-[M_s]=\epsilon/2-\beta/\nu$~\cite{zhong24}. The second one is the noise amplitude. It scales as $\sigma b^{[\sigma]}$ with $[\sigma]=z+2\beta/\nu$~\cite{zhong24}. Both can be directly obtained from a dimension analysis to Eq.~(\ref{lang2d}). Taking into account these two explicit factors and setting $b=\sigma^{-1/[\sigma]}$, we arrive at
\begin{equation}
m=\sigma^{\frac{\beta}{\nu[\sigma]}}\tilde{g}(\tau \sigma^{-\frac{1}{\nu[\sigma]}}, h\sigma^{-\frac{\beta\delta}{\nu[\sigma]}}, (t-t_0)\sigma^{\frac{z}{[\sigma]}},a_4\sigma^{-\frac{[a_4]}{[\sigma]}}),\label{mht}
\end{equation}
where $\tilde{g}$ is another scaling function and we have replaced $a_3$ with $a_4$ since $a_3=3a_4M_s$. We have also introduced an initial time $t_0$ to which we will return later on. Moreover, corrections to scaling including deviation from the fixed point have been neglected~\cite{Wegner,zhong16,Justin,Amit}.

In the MF theory, valid above $d>6$, we can simply fix the arguments of $\tilde{g}$ by choosing different $\tau$ (through $a_2$), $a_4$, and $H$ for different $\sigma$ according to Eq.~(\ref{mht}) and verify the theory through $M\sigma^{-\beta/\nu[\sigma]}$ versus $t\sigma^{z/[\sigma]}$ ($t_0=0$) curve collapse. In this way, we do not need to subtract $M_s$ and $H_s$ because they just scale identically with $M$ and $H$, respectively, and thus contribute an overall displacement of the collapsed curves. This is the reason why we do not need to know them.

However, when fluctuations are taken into account in $d<6$, further complications emerge~\cite{zhong24}. We note from Eq.~(\ref{da2}) that $[\delta a_2]=[a_4]+d-2=(d+2)/2-\beta/\nu$, which is different from $[\delta\tau]=2[a_3]+d-4=2=[\tau]=[a_2]$, using Eq.~(\ref{dth}), except in MF. So is $[a_4M_s^2]=\epsilon/2+\beta/\nu$. These indicate that the two quartic contributions to $\tau$, Eq.~(\ref{tauh}), scales differently from both $a_2$ itself and $\tau$ and $\delta\tau$, unlike the MF case. Similarly, according to Eq.~(\ref{dth}), $[\delta h]=[a_3]+d-2=(d+2)/2=[h]$, whereas $[a_2M_s]=2+\beta/\nu$ and $[a_4M_s^3]=\epsilon/2+2\beta/\nu$. These mean that each term of the quartic contributions to $H_s$, $\hat{H}_s$, Eq.~(\ref{tauh}), scales differently from $h$ and $\delta h$ except again in MF. Owing to these singular contributions with different dimensions from the quartic theory, the MF method above can no longer fix the corresponding scaling arguments.

The solutions to these two problems are firstly to choose $a_2$ as $\tau-\delta a_2-3a_4M_s^2$ such that $a_2+\delta a_2+3a_4M_s^2=\tau$ and secondly to subtract $\hat{H}_s$, instead of the entire $H_s$, out of $H$~\cite{zhong24}. Note that we have explicitly written out the singular contribution $\delta a_2$ to $a_2$ in $\tau$, Eq.~(\ref{tauh}). For a reference curve run with parameters $a_{20}$, $a_{40}$, $H_0$, and $\sigma_0$, To keep $\tau \sigma^{-1/\nu[\sigma]}=\tau_0\sigma_0^{-1/\nu[\sigma]}$ and $(H-\hat{H}_s)\sigma^{-\beta\delta/\nu[\sigma]}=(H_0-\hat{H}_{s0})\sigma_0^{-\beta\delta/\nu[\sigma]}$ (besides a similar relation for $a_4$) from Eq.~(\ref{mht}), we thus choose
\begin{eqnarray}
a_2&=&\left\{a_{20}+\delta \hat{a}_2+3a_{40}M_{s0}^2\left[1-(\sigma_0/\sigma)^{\rho}\right]\right\}(\sigma_0/\sigma)^{-1/\nu[\sigma]}\label{a2}\nonumber\\
H&=&\hat{H}_s+(H_0-\hat{H}_{s0})(\sigma_0/\sigma)^{-\beta\delta/\nu[\sigma]}
\end{eqnarray}
with $\delta \hat{a}_2=\delta a_{20}-\delta a_2(\sigma_0/\sigma)^{1/\nu[\sigma]}$ and $\rho=1/\nu[\sigma]-[a_4M_s^2]/[\sigma]$ by adjusting $M_{s0}$ and $\delta\hat{a}_2$ so that the new curve overlaps the reference one after being rescaled according to Eq.~(\ref{mht}) (with $h$ replaced by $H-\hat{H}_{s0}$ and $m$ by $M$).

To test the theory, we solve Eq.~(\ref{lang2d}) by direct Euler discretization. In the zero-dimensional case, the time step is $0.0005$ for the small noises but $0.0002$ for large ones, both having been checked to bring stable results. The average is over $200~000$ samples. The same discretization is also applied to a two-dimensional system. The space step is fixed to 1, while the time step is $0.01$, again checked to be sufficient. The lattice is $200\times200$ with periodic boundary conditions. More than $10000$ samples are employed for average.

\begin{figure}
\centerline{\includegraphics[width=\linewidth]{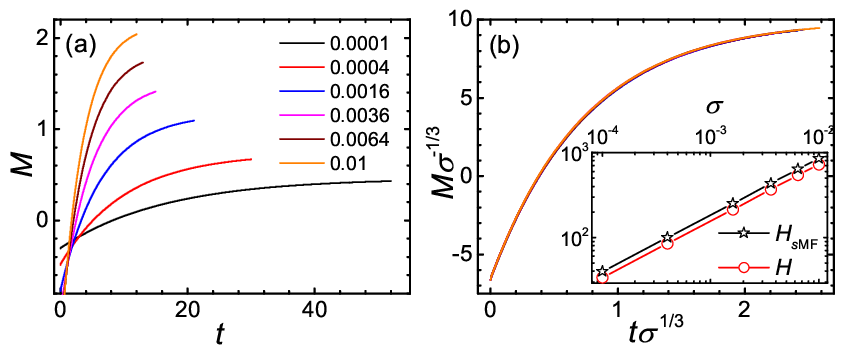}}
\caption{\label{tm0d}(Color online)
(a) Temporal evolution of the order parameter $M$ for a series of the noise amplitude $\sigma$ listed from numerical solutions of Eq.~(\ref{lang2d}) in the absence of the gradient term. $a_2=a_{20}(\sigma_0/\sigma)^{-1/3}$, $a_4=a_{40}(\sigma_0/\sigma)^{1/3}$, and $H=H_0(\sigma_0/\sigma)^{-2/3}$ with $a_{20}=-400$, $a_{40}=948$, $H_0=85$, $\sigma_0=0.0004$, and $\lambda=0.0005$. (b) Rescaling of all curves in (a) according to Eq.~(\ref{mht}). The inset displays $H$ and $H_{s{\rm MF}}$ versus $\sigma$. Lines connecting symbols are only a guide to the eye.}
\end{figure}
We first consider the zero-dimensional case, which, as mentioned, has been shown to control by the MF exponents even in the presence of the noise~\cite{zhong24}. One sees that all curves in Fig.~\ref{tm0d}(a) perfectly collapse onto a single curves in Fig.~\ref{tm0d}(b) in complete compliance with Eq.~(\ref{mht}), since all arguments but the one containing $t$ in $\tilde{g}$ are kept constant. Any state with an initial negative order parameter converges rapidly to the curves displayed and $t_0=0$. The evolution curves in Fig.~\ref{tm0d}(a) result from activated barrier crossing from $-M_{\rm eq}$ to $+M_{\rm eq}$ since the applied field $H$ is smaller than the MF spinodal field as is evident from the inset in Fig.~\ref{tm0d}(b). This is similar to nucleation and growth though the curves exhibit no incubation, a plateau of metastable states, even for the smallest $\sigma$. The reason is that the noise is Gaussian and sufficient samples can always achieve barrier crossing. Note that one can also employ $H$, say, to rescale the curves because $H\sigma^{-2/3}$ is fixed.

\begin{figure*}
\centerline{\includegraphics[width=\linewidth]{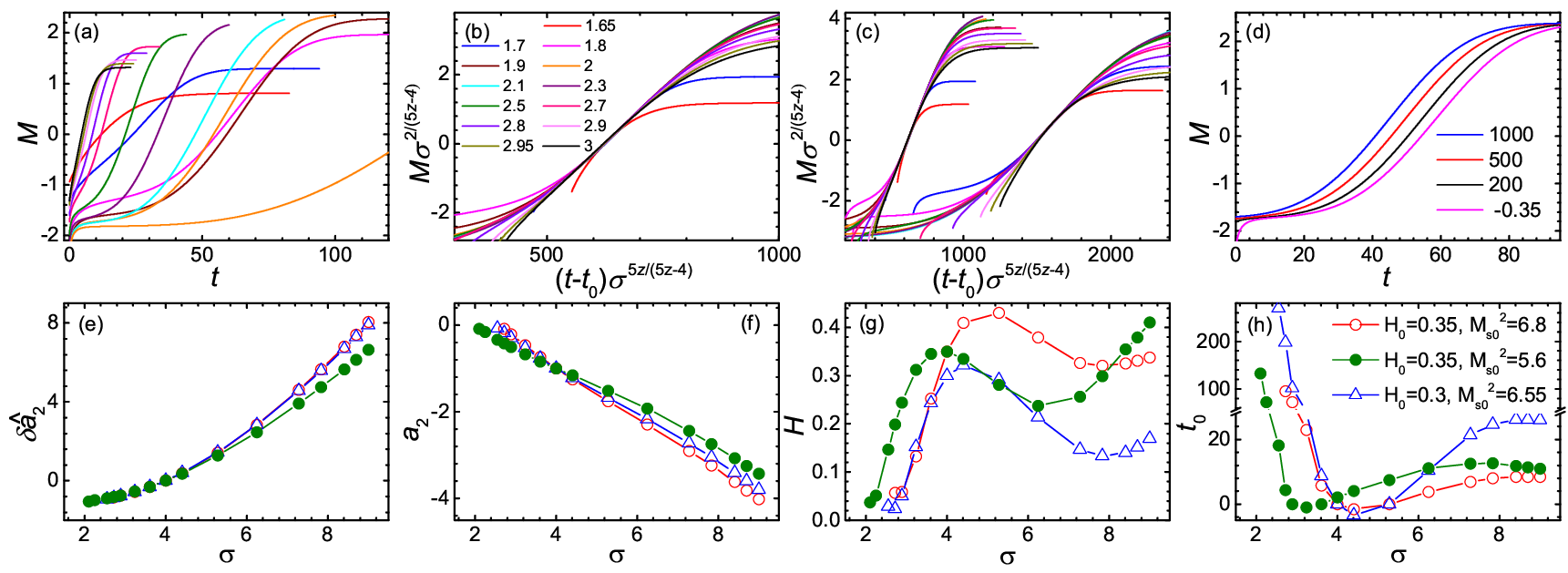}}
\caption{\label{tm2d}(Color online)
(a) Temporal evolution of $M$ for a series of $\sqrt{\sigma}$ given in (b) from numerical solutions of Eq.~(\ref{lang2d}) in two dimensions. $\lambda=1$, $a_{20}=-1$, $a_{40}=1/6$, $H_0=0.35$, and $\sigma_0=2^2$ for the reference curve. $a_4=a_{40}(\sigma_0/\sigma)^{-12/(5z-4)}$, $M_s=M_{s0}(\sigma_0/\sigma)^{2/(5z-4)}$ with $M_{s0}^2=6.8$, $z=1.85$, and $a_2$ and $\hat{H}_s$ given by Eq.~(\ref{a2}) for $\delta\hat{a}_2$ plotted in (e) for other curves. The lowest curve is another reference curve with another $H_0=0.3$. The evolution curves move from top left to right and then back to the left again as $\sigma$ decreases from $3^2$ to $1.9^2$ and further decreases to $1.65^2$. (b) Rescaling of all curves but the lowest one in (a) according to Eq.~(\ref{mht}). The curve collapses are good except for crossovers (see the text). (c) Another view of the collapsed curve in (b) (left) and another collapses for $H_0=0.3$ and $M_{s0}=6.55$ (right). (d) Evolution curves with the reference parameters of $H_0=0.35$ evolving from an identical initial state but subject to different number of initial equilibration steps listed in the legend. The rightmost curve equilibrates at $H_0=-0.35$ for $1000$ steps before the field is flipped at $t=0$. (e)--(h), $\sigma$ dependence of $\delta\hat{a}_2$, $a_2$, $H$, and $t_0$ for $H_0=0.35$ and $M_{s0}^2=6.8$ (open circles) and $M_{s0}^2=5.6$ (filled circles) and $H_0=0.3$ and $M_{s0}^2=6.55$ (triangles). Lines connecting symbols are only a guide to the eye.}
\end{figure*}
Next we study the two-dimensional case in which $\nu=-5/2$, $\delta=-6$~\cite{Cardy85}, and $\beta=1$ exactly and $z=1.85$~\cite{zhong24}. For the reference curve with $H_0=0.35$ and $\sigma_0=2^2$, an incubation stage in which sufficiently large nuclei of the equilibrium phase needs to be nucleated from the metastable state before growing to cover the whole system is evident from Fig.~\ref{tm2d}(a). We find that $M_{s0}^2=6.8$ and $\delta\hat{a}_2=1.431$ can collapse the curve of $\sigma=2.3^2$ with the reference curve except for the metastable and equilibrium states as expected. For other values of $\sigma$, the same $M_{s0}$ requires unique values of $\delta\hat{a}_2$ to render the corresponding curves parallel to the reference curve. An overall displacement of $t$ by a unique $t_0$ is indispensable for the curve collapse. The results are presented in Fig.~\ref{tm2d}(b) and the corresponding parameters are depicted in Figs.~\ref{tm2d}(e)--\ref{tm2d}(h), which demonstrate the smooth variations of the parameters with $\sigma$, although we separately adjust $\delta \hat{a}_2$ and $t_0$ for each curve, thereby showing the consistency of the results. We have also shown in Fig.~\ref{tm2d}(c) the collapse of another reference curve with a smaller $H_0=0.3$ that has a much longer incubation stage as seen from the lowest curve in Fig.~\ref{tm2d}(a). We note that all $H$ values are smaller than the corresponding $H_{s{\rm MF}}$.

The salient difference to the driving case is that we need $t_0$ here. To understand why we need it, we plot in Fig.~\ref{tm2d}(d) evolution curves with the reference parameters and starting with an identical initial state but subject to different stages of initial equilibration. It is seen that as the number of initial equilibration steps decreases, the curves move to the right since more further steps are needed to transition to the equilibrium state. This is in stark contrast to the driving case in which the position of a driving curve exhibits little dependence on the initial conditions once the initial field is sufficiently far away from the transition~\cite{zhong24}. Although, for uniformity, we all adopt the same initiation of equilibration at $-H$ and flipping the field at $t=0$, it can be seen from Fig.~\ref{tm2d}(a) that the transient stage from $t=0$ to the metastable state is markedly different. This different initial condition is thus accounted for by $t_0$.

There is yet another origin of $t_0$. We can choose $M_{s0}^2=5.6$ to collapse the curve of $\sigma=1.8^2$ with the same reference curve with $t_0=0$. This results in another set of parameters with nonlinear relationship, also displayed in Figs.~\ref{tm2d}(e)--\ref{tm2d}(h). The collapsed curve is the same as that of $M_{s0}^2=6.8$ with the only difference being that the coincidence range of each $\sigma$ curve is different and is thus not shown. Therefore, $t_0$ also bears the differences both in the contribution of $M_{s0}^2$ to $a_2$ and in $\delta\hat{a}_2$ in Eq.~(\ref{a2}). A different choice of $M_{s0}$ then also implies the choice of a different $t_0$. In addition, lattice size effects are also accounted for in this way. One might then worry about the definiteness of the results as there exist up to three adjustable parameters. However, we emphasize that given a $M_{s0}$, $\delta\hat{a}_2$ and $t_0$ are uniquely determined. The freedom in the selection of $M_{s0}$ reflects the existence of the scaling and the effectiveness of the method.

We note that the curve collapses in Figs.~\ref{tm2d}(b) and~\ref{tm2d}(c) are really good even though they are different from that in Fig.~\ref{tm0d}(b). This difference is expected as can be clearly seen from the middle red curve with $\sigma=1.65^2$. Its equilibrium states do not overlap those of other noises and its size is much smaller than others. This is expectable as the equilibrium and metastable states alike are not described by the cubic fixed point and its exponents. Excluding these two kinds of states and their crossovers to the intermediate states that are controlled by the cubic fixed point, these intermediate states occupy a large portion of the entire transitional region and well collapse onto those similar states of other noises, forming an envelope similar to the collapsed curve in Fig.~\ref{tm0d}(b). In addition, the range of $\sigma$ for the curve collapse shown is not as large as that in Fig.~\ref{tm0d}. This is limited by $\sigma$. Too small a $\sigma$ has no transition at all since $a_2>0$ from Fig.~\ref{tm2d}(f) while too large a $\sigma$ makes the transition occur rapidly though its $H$ is much smaller than its $H_{s{\rm MF}}$. The range also depends on $M_{s0}$. The smaller the $M_{s0}$, the larger the range, as seen in Figs.~\ref{tm2d}(g) and~\ref{tm2d}(h). However, the range of $H$ in the present $\sigma$ range covers one order of magnitude and up to over $17$ times for $M_{s0}^2=6.8$, see Fig.~\ref{tm2d}(g). This is in fact a large range for nucleation as $H$ appears as a driving force for nucleation in an exponential~\cite{zhong18}. For example, the two reference curves in Fig.~\ref{tm2d}(a) already separate a lot though their $H$ values change little. Furthermore, the field range is just over $21$ times in Fig.~\ref{tm0d}. Therefore, unlike their appearance, the curve collapses are in fact really good and the corresponding field range is relatively large for nucleation.

In conclusion, we have generalized the method developed recently for the complete universal scaling of driving FOPTs to the case of standard nucleation and growth. A new nonuniversal initial time must be introduced to collapse different evolution curves. It accounts for the differences both in the initial evolution time and in the effective temperatures and thus the field strengths. We have clearly demonstrated that through subtracting non-universal contributions such as $\delta\hat{a}_2$, $\hat{H}_s$, as well as $t_0$, universal scaling in nucleation and growth does emerge. This has been achieved by assuming the existence of the spinodal points to reach the cubic theory and by carefully considering the singular fluctuation contributions of the quartic theory to the cubic one. It is remarkable that such a straightforward but systematic approach can reveal the universal scaling behavior in FOPTs both through driving and through standard nucleation and growth. As FOPTs exist far more widespread in nature compared to continuous phase transitions, applications of the present theory and method are thus highly anticipated. This may also shed light to anomalous phase transitions such as deconfined quantum critical points~\cite{senthil} in which FOPTs may be a possible alternative.

\begin{acknowledgments}
This work was supported by National Natural Science Foundation of China (Grant No. 12175316).
\end{acknowledgments}


\begin{thebibliography}{99}


\bibitem{Becker}M. Volmer and A. Weber, Nucleus formation in supersaturated systems, Z. Phys. Chem. (Leipzig) {\bf 119}, 277-301 (1926).
\bibitem{Becker1}L. Farkas, Keimbildungsgeschwindigkeit in \"{u}bers\"{a}ttigten D\"{a}mpfen. {\it ibid.} {\bf 125,} 236-242 (1927).
\bibitem{Becker2}R. Becker and W. D\"{o}ring, Kinetic treatment of grain formation in supersaturated vapours, Ann. Phys. (Leipzig) {\bf 24}, 719-752 (1935).
\bibitem{Zeldovich}Ya. B. Zeldovich, On the theory of new phase formation: cavitation, Acta Physicochim. USSR {\bf 18}, 1-22 (1943).

\bibitem{Avrami} A. N. Kolmogorov, On the statistical theory of the crystallization of metals, Bull. Acad. Sci. USSR (Sci. Mater. Nat.) {\bf 3}, 355-359 (1937)
\bibitem{Avrami1}W. A. Johnson and P. A. Mehl, Reaction kinetics in processes of nucleation and growth, Trans. Am. Inst. Min. Metall. Eng. {\bf 135}, 416-442 (1939).
\bibitem{Avrami2}M. Avrami, Kinetics of phase change. I General theory, J. Chem. Phys. {\bf 7}, 1103-1112 (1939).
251 (1990).
\bibitem{Langer67}J. S. Langer, Theory of the condensation point, Ann. Phys. (N. Y.) {\bf 41}, 108 (1967).
\bibitem{books}F. F. Abraham, {\it Homogeneous Nucleation Theory} (Academic, New York, 1974).
\bibitem{books1} P. Debenedetti, {\it Metastable Liquids} (Princeton University Press, Princeton, NJ 1996).
\bibitem{books2} D. Kashchiev, {\it Nucleation: Basic Theory with Applications} (Butterworth-Heinemann, Oxford, 2000).

\bibitem{Oxtoby92}D. W. Oxtoby, Nucleation of first-order phase transitions, Acc. Chem. Res. {\bf 31}, 91-97 (1998).
\bibitem{Oxtoby921}J. D. Gunton, Homogeneous nucleation, J. Stat. Phys. {\bf 95}, 903-923 (1999).
\bibitem{Oxtoby922}S. Auer and D. Frenkel, Quantitative prediction of crystal-nucleation rates for spherical colloids: A computational approach, Annu. Rev. Phys. Chem. {\bf 55}, 333-361 (2004).
\bibitem{Oxtoby923}R. P. Sear, Nucleation: theory and applications to protein solutions and colloidal suspensions, J. Phys.: Condens. Matter {\bf 19}, 033101 (2007).
\bibitem{Binder16} K. Binder and P. Virnau, Overview: Understanding nucleation phenomena from simulations of lattice gas models, J. Chem. Phys. {\bf 145}, 211701 (2016).


\bibitem{zhong18} F. Zhong, Universal scaling in first-order phase transitions mixed with nucleation and growth, J. Phys. Condens. Matters \textbf{30}, 445401 (2018).

\bibitem{Gunton83}J. D. Gunton, M. San Miguel, and P. S. Sahni, The dynamics of first-order phase transitions, in {\it Phase Transitions and Critical Phenomena}, eds. C. Domb and J. L. Lebowitz Vol. \textbf{8}, 267 (Academic, London, 1983).
\bibitem{Binder} K. Binder, Theory of first-order phase transitions, Rep. Prog. Phys. {\bf 50}, 783 (1987).
\bibitem{Binder2} K. Binder and P. Fratzl, Spinodal decomposition. in Phase Transformations in Materials, ed. Kostorz, G. 409-480 (Wiley, Weinheim, 2001).

\bibitem{Stanley} H. E. Stanley, {\it Introduction to Phase Transitions and Critical Phenomena} (Oxford University, Oxford, 1971).
\bibitem{Mask}S.-k. Ma, {\it Modern Theory of Critical Phenomena} (W. A. Benjamin Inc., Canada, 1976).
\bibitem{Cardyb}J. Cardy, {\it Scaling and Renormalization in Statistical Physics} (Cambridge University Press, Cambridge, 1996).
\bibitem {Justin}J. Zinn-Justin, \textit{Quantum Field Theory and Critical Phenomena}, 5th edition (Oxford University Press, Oxford, 2021).
\bibitem{Amit}D. J. Amit and V. Martin-Mayer, {\it Field Theory, the Renormalization Group, and Critical Phenomena}, 3rd edition (World Scientific, Singapore, 2005).
\bibitem{zhong24} F. Zhong, Complete universal scaling in first-order phase transitions, arXiv:2404.00219 (2024).
\bibitem{Filion}S. Auer and D. Frenkel, Prediction of absolute crystal-nucleation rate in hard-sphere colloids, Nature (London) {\bf 409}, 1020-1023 (2001).
\bibitem{Filion1}T. Kawasaki and H. Tanaka, Formation of a crystal nucleus from liquid, Proc. Nat. Acad. Sci. {\bf 107}, 14036-14041 (2010).
\bibitem{Filion2}L. Filion, R. Ni, D. Frenkel, and M. Dijkstra, Simulation of nucleation in almost hard-sphere colloids: The discrepancy between experiment and simulation persists, J. Chem. Phys. {\bf 134}, 134901 (2011).
\bibitem{zhongl05}F. Zhong and Q. Z. Chen, Theory of the dynamics of first-order phase transitions: Unstable fixed points, exponents, and dynamical scaling, Phys. Rev. Lett. \textbf{95}, 175701 (2005).

\bibitem{zhong16}F. Zhong, Renormalization-group theory of first-order phase transition dynamics in field-driven scalar model, Front. Phys. \textbf{12}, 126402 (2017).
\bibitem{zhong12} F. Zhong, Imaginary fixed points can be physical, Phys. Rev. E \textbf{86}, 022104 (2012).

\bibitem{Wegner} F. W. Wegner, Corrections to scaling laws, Phys. Rev. B {\bf 5,} 4529-4536 (1972).


\bibitem{Cardy85}J. L. Cardy, Conformal invariance and the Yang-Lee edge singularity in two dimensions, Phys. Rev. Lett. {\bf 54}, 1354 (1985).
\bibitem{senthil} T. Senthil, A. Vishwanath, L. Balents, S. Sachdev, and M. P. A. Fisher, Deconfined Quantum Critical Points, Science {\bf 303}, 1490 (2004).

\end{thebibliography}
\end{document}